\pdfoutput=1

\documentclass[11pt]{article}

\usepackage[preprint]{acl}

\usepackage{times}
\usepackage{latexsym}
\usepackage[T1]{fontenc}
\usepackage[utf8]{inputenc}
\usepackage{microtype}
\usepackage{inconsolata}
\usepackage{graphicx}
\usepackage{xcolor}
\usepackage{url}
\usepackage{booktabs}
\usepackage{multirow}
\usepackage{svg}
\usepackage{tikz}
\usepackage{hyperref}
\usepackage{amsmath,amsfonts,amssymb}
\usepackage{nicefrac}
\usepackage{todonotes}
\usepackage{wrapfig}
\usepackage{subcaption}
\usepackage{colortbl}
\usepackage{pgfplots}
\usepackage{pgfplotstable}
\pgfplotsset{compat=1.18}

\definecolor{myblue}{rgb}{.1, 0.7, 1}
\definecolor{darkorange}{rgb}{.9, .5, .1}
\definecolor{darkgreen}{rgb}{0, .4, .13}
\definecolor{lightblue}{rgb}{.8, 1, 1}
\definecolor{lightorange}{rgb}{.9, .5, .1}

\newcolumntype{g}{>{\columncolor{lightblue}}c}

\newcommand{\readapt}{{\sc RE-Adapt}}
\newcommand{\readaptir}{{\sc RE-AdaptIR}}

\title{RE-AdaptIR: Improving Information Retrieval\\ through Reverse Engineered Adaptation}

\author{%
William Fleshman \and Benjamin Van Durme \\
Johns Hopkins University \\
\texttt{will.fleshman@jhu.edu}}
  
\begin{document}
\maketitle
\begin{abstract}
Large language models (LLMs) fine-tuned for text-retrieval have demonstrated state-of-the-art results across several information retrieval (IR) benchmarks. However, supervised training for improving these models requires numerous labeled examples, which are generally unavailable or expensive to acquire. In this work, we explore the effectiveness of extending reverse engineered adaptation to the context of information retrieval (RE-AdaptIR). We use RE-AdaptIR to improve LLM-based IR models using only unlabeled data. We demonstrate improved performance both in training domains as well as zero-shot in domains where the models have seen no queries. We analyze performance changes in various fine-tuning scenarios and offer findings of immediate use to practitioners.
\end{abstract}

\section{Introduction}

Information retrieval (IR) is a fundamental component of various modern applications, powering search engines, recommender systems, and various data analytics pipelines. Recently, large language models (LLMs) have achieved state-of-the art results on dense text retrieval, identifying and ranking the most relevant text for a given query by comparing learned vector representations of the text \citep{reimers-gurevych-2019-sentence, colbert, karpukhin-etal-2020-dense,izacard2022unsupervised,ma2023finetuning,jiang2023mistral,weller2024followir}. The effectiveness of text retrieval systems have a direct impact on numerous domains, including healthcare, finance, and social media, where accurate and timely access to information is critical. Retrieval is also critical in the context of retrieval augmented generation (RAG), enabling LLMs access to external resources when constructing a response \citep{rag}. For these reasons, we seek a practical and efficient approach for improving existing text retrieval models.

\begin{figure}[h]
    \centering
    \includegraphics[width=\columnwidth]{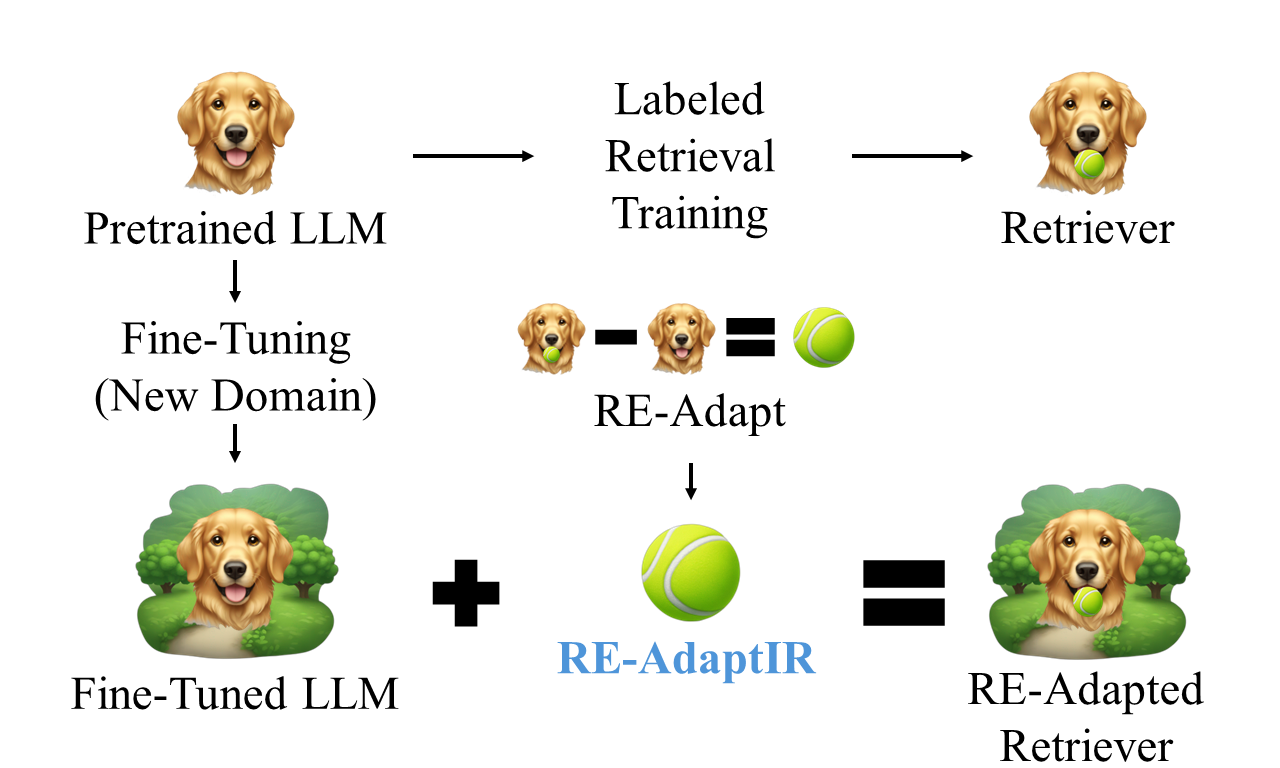}
    \caption{In \readaptir{}, \readapt{} is extended to an existing retrieval model to isolate what was learned during labeled contrastive training. The pretrained model is fine-tuned on unlabeled in-domain documents and \textit{readapted} for text retrieval. The new retriever outperforms the original on both in-domain and zero-shot information retrieval tasks.}
    \label{fig:cartoon}
\end{figure}

Supervised fine-tuning of LLMs for text retrieval tasks has become a widely adopted approach, leveraging their pretrained language understanding capabilities to achieve state-of-the-art results on various benchmarks \citep{ma2023finetuning,jiang2023mistral,weller2024followir}. However, adapting an LLM for text retrieval requires labeled datasets, with numerous example queries and documents both related and unrelated to forming a helpful response. This poses a significant challenge to improving these systems, as data annotation or synthetic generation can be too expensive, difficult, and error-prone \citep{datalabel, pmlr-v133-desmond21a}. Making matters worse, fine-tuning an existing LLM on new domains can cause \textit{forgetting}, a decreased performance on previously capable tasks \citep{MCCLOSKEY1989109,kotha2024understanding}. \citet{fleshman2024readapt} recently proposed reverse engineered adaptation (\readapt), an approach for solving similar dilemmas faced when fine-tuning existing instruction-tuned models. Here we introduce \readaptir, an extension of \readapt{} to IR models, leveraging unlabeled data to improve existing text-retrieval LLMs (\autoref{fig:cartoon}). Specifically we:

\begin{itemize}
    \item Extend \readapt{} to the information retrieval setting and apply \readaptir{} to two state-of-the-art text retrieval models: RepLLaMA and e5-Mistral;
    \item Demonstrate improved performance both in-domain and zero-shot across 14 datasets; and
    \item Explore the importance of fine-tuning on data relevant to test-time queries and the impact different scenarios have on performance.
\end{itemize}

\section{Background}

\subsection{Retrieval Models}

The transformer architecture is a natural choice for text retrieval models, as it embeds text into dense vector representations \citep{vaswani2023attention}. LLMs pretrained on massive amounts of text have demonstrated the ability to capture useful semantic meaning in these vector representations \citep{Radford2019LanguageMA,touvron2023llama, jiang2023mistral}. To ensure these models are capable for retrieval, a secondary fine-tuning stage is used to encourage the model to map similar texts to similar vectors \citep{ma2023finetuning,jiang2023mistral,weller2024followir}. This is generally done via some form of supervised contrastive training such as with InfoNCE \citep{oord2019representation}. 
After training, the models can be used to create a vector database of documents, and rank them given a query representation by using a similarity function. Related documents should have more similar representations to the query than those unrelated. We experiment with two such models in this work: RepLLaMA \citep{ma2023finetuning} and e5-Mistral \citep{wang2024improving}. 

\paragraph{RepLLaMA.} \citet{ma2023finetuning} introduced RepLLaMA to demonstrate that state-of-the-art LLMs could surpass the previous results achieved with smaller retrieval models, especially when evaluated zero-shot on datasets not seen during training. They construct RepLLaMA by fine-tuning LLaMA-2-7B \citep{touvron2023llama} on approximately 500k labeled examples from the training split of the MS-MARCO dataset \citep{bajaj2018ms}. 

\paragraph{e5-Mistral.} In a similar fashion, \citet{wang2024improving} fine-tune the Mistral-7B LLM \citep{jiang2023mistral} using a combination of synthetic data, MS-MARCO, and multiple other labeled datasets. The resulting e5-Mistral model achieves state-of-the-art results on several text retrieval benchmarks. 

In both cases, labeled data was needed to achieve the best results, and it is unclear how to effectively incorporate the copious amount of unlabeled text additionally available. For example, MS-MARCO contains almost 9 million passages, but only a fraction of this data is used in training, due to the limited number of associated queries available. In this work, we use \readaptir{} to leverage this unlabeled data and improve the performance of both RepLLaMA and e5-Mistral.

\subsection{Reverse Engineered Adaptation}

\citet{fleshman2024readapt} introduced reverse engineered adaptation (\readapt) as a new method to efficiently update instruction-tuned models with unlabeled data lacking the previously required instruction-tuning annotations \citep{mishra-etal-2022-cross,wei2022finetuned,Ouyang2022TrainingLM}. \readapt{} works by first isolating what has been learned from instruction-tuning by taking the difference between the weights of the instruction-tuned and pretrained versions of a model. This difference can be thought of as an adapter \cite{pmlr-v97-houlsby19a} or as a multi-task version of task-vectors \citep{ilharco2023editing}. Given this RE-Adapter $\Delta$, the pretrained weights $\Theta$ can be fine-tuned with a new \textit{knowledge adapter} $\Psi$ without impacting the previous instruction-tuning. Finally, the model can be re-instantiated with weights $\Theta + \alpha \Psi + \beta \Delta$ where $\alpha$ and $\beta$ are \textit{partial adaptation} scalars used to control the strength of fine-tuning \citep{fleshman2024readapt}. The authors show that \readapt{} improves the performance of instruction-tuned models in the new domain while preserving or improving performance out-of-domain. 

\section{Re-AdaptIR}

In this work, we explore the effectiveness of \readapt{} in the context of information retrieval. While an instruction-tuned model still leverages the pretraining capabilities of next-token prediction, most text retrieval models do not. RepLLaMA and e5-Mistral both discard the next-token predictor from their respective LLMs and fine-tune the model to produce a single vector representation per document \citep{ma2023finetuning, wang2024improving}. It is therefore unclear whether continued fine-tuning of the pretrained LLM with next-token prediction will improve down-stream retrieval. 

To answer this question, we first fine-tune the pretrained LLM on unlabeled documents from a new domain. We then construct a RE-AdaptIR for the retrieval model by discarding the pretrained next-token predictor weights as well as the corresponding weights from the knowledge adapter. We can then follow the \readapt{} procedure using the remaining weights, and evaluate the \textit{readapted} IR model using queries from the new domain.

\section{Experiments}

We first replicate the shared text retrieval experiments conducted by \citet{ma2023finetuning} and \citet{wang2024improving} and compare the base model performance before and after applying \readaptir. We then conduct further analysis to understand how different fine-tuning scenarios impact performance. 

\subsection{Datasets}

Both RepLLaMA and e5-Mistral utilized MS-MARCO \citep{bajaj2018ms} as part of their training data, and we include it in our evaluations to help measure any benefits from using \readaptir{} in-domain. Specifically, \readaptir{} allows for fine-tuning over the entire 8.84M passages, where only a subset of those passages were used for retrieval training due to limited availability of query-passage pairs \citep{ma2023finetuning, wang2024improving}. Additionally, we use the same 13 public datasets from the BeIR IR benchmark \citep{thakur2021beir} used by \citet{ma2023finetuning} to assess RepLLaMA's zero-shot performance across a diverse set of IR tasks. Of these, we note that the training splits of FEVER \citep{thorne-etal-2018-fever}, HotPotQA \citep{yang-etal-2018-hotpotqa}, NQ \citep{kwiatkowski-etal-2019-natural}, and Quora \citep{quora-question-pairs} were also used by \citet{wang2024improving} in the training of e5-Mistral, providing more in-domain insight to our experiments. These datasets are zero-shot for RepLLaMA, as is the remainder of BeIR for both models. We use the same prompts used by \citet{ma2023finetuning} and \citet{wang2024improving} for each dataset.

\subsection{Adapters}

We use parameter efficient fine-tuning \citep{peft} with DoRA \citep{liu2024dora} to adapt the pretrained LLaMA-2-7B \citep{touvron2023llama} and Mistral-7B \citep{jiang2023mistral} LLMs. We train each adapter for a single epoch on all passages from the dataset under evaluation. Specific training details are included in \autoref{sec:details}. As in \citet{fleshman2024readapt}, we use a scalar of 0.5 with our knowledge adapters to minimize interference with existing retrieval ability.

\begin{table}[ht]
    \centering
    \small
    \begin{tabular}{l|cg|cg}
        & \multicolumn{2}{c|}{\textbf{RepLLaMA}} & \multicolumn{2}{c}{\textbf{e5-Mistral}}\\
        \textbf{Dataset} & \textbf{Base} & \textbf{RA} & \textbf{Base} & \textbf{RA} \\
        \toprule
        MS-MARCO & \cellcolor{orange!25}\textbf{46.5} & 46.1 & \cellcolor{orange!25} 36.5 & \textbf{40.1}\\
        FEVER & \textbf{84.0} & 83.8 & \cellcolor{orange!25} 85.1 & \textbf{87.7}\\
        HotPotQA & 67.2 & \textbf{67.6} & \cellcolor{orange!25}72.5 & \textbf{73.4}\\
        NQ & 61.8 & \textbf{62.1} & \cellcolor{orange!25} \textbf{53.3} & 52.4\\
        Quora & 80.1 & \textbf{82.8} & \cellcolor{orange!25}85.6 & \textbf{88.0} \\
        Arguana & 52.3 & \textbf{52.6} & 52.0 & \textbf{59.0} \\
        Climate-FEVER & \textbf{30.8} & 30.4 & 24.9 & \textbf{31.4}\\
        DBPedia & 43.4 & \textbf{43.5} & \textbf{47.2} & 47.1\\
        FiQA & 44.2 & \textbf{45.5} & 49.9 & \textbf{52.3}\\
        NFCorpus & 38.0 & \textbf{38.6} & 39.6 & \textbf{40.8} \\
        SCIDOCS & 17.7 & \textbf{18.3} & 18.6 & \textbf{18.7} \\
        SciFact & 74.5 & \textbf{76.3} & 71.4 & \textbf{73.3} \\
        TREC-COVID & 84.0 & \textbf{85.6} & \textbf{83.9} & 81.0 \\
        Touche-2020 & \textbf{27.5} & 27.0 & 29.0 & \textbf{30.1}\\
        \midrule
        \textbf{Average} & 53.7 & \textbf{54.3} & 53.5 & \textbf{55.4}\\
        \textbf{Average Z-Shot} & 54.3 &  \textbf{54.9} & 46.3 & \textbf{48.2}\\
    \end{tabular}
    \caption{nDCG@10 across test splits for MS-MARCO and BeIR datasets. The results highlighted in orange indicate the dataset's train split was used for training the corresponding model and are not zero-shot. \textit{Base} is the unmodified model and \textit{RA} is the model RE-Adapted after fine-tuning on the domain.}
    \label{tab:main}
\end{table}

\subsection{Results}

Our main results are compiled in \autoref{tab:main}. We see that \readaptir{} improves results in the majority of cases, increasing the average zero-shot nDCG@10 by 0.6 and 1.9 points for RepLLaMA and e5-Mistral respectively. \textbf{Importantly, these performance gains required no additional labeled data} and are achieved by simply fine-tuning the pretrained model over the document database being used for retrieval. The default partial adaptation scalar of 0.5 was used for this experiment, but we note that optimizing this value per dataset does improve results. While not applicable to our zero-shot analysis, in practice this value can be set using withheld queries. 

\begin{figure}[h]
    \centering 
    \begin{tikzpicture}
    \begin{axis}[width=.7\columnwidth,
          grid=major,
          grid style={dashed,gray!30},
          ylabel=$\Delta$nDCG@10,
          xlabel=Corpus Size,
          xmode=log,
          legend style={legend pos=north east,font=\small}
          ]
        \addplot[only marks, mark size=4pt, myblue, opacity=0.5, mark=diamond*] table[x={size}, y={change}, col sep=comma]{data/llama_size.csv};
        \addlegendentry{RepLLaMA};
        
        \addplot[only marks, mark size=3pt, darkorange, opacity=0.5, mark=square*] table[x={size}, y={change}, col sep=comma]{data/mistral_size.csv};
        \addlegendentry{e5-Mistral};
        \addplot[line width=3pt, black, forget plot] table[x={size}, y={create col/linear regression={y={change}}}, col sep=comma]{data/combined_size.csv};
    \end{axis}
    \end{tikzpicture}
    \caption{The observed relationship between the corpus size and the change in performance when fine-tuning with \readaptir{}.}
    \label{fig:size}
\end{figure}
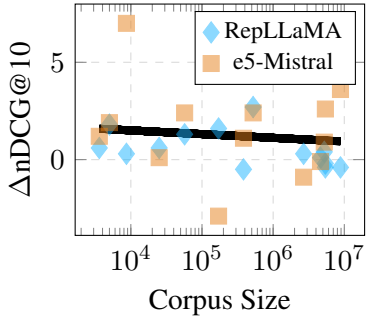

We notice the few cases where performance was reduced tend to occur with the larger corpora. We plot this relationship in \autoref{fig:size} and do observe a slightly negative correlation. This relationship is purely observational and likely caused by latent topic or task diversity among the larger datasets used in these experiments. For reference, we include the dataset sizes in \autoref{sec:sizes}.

\begin{table*}[!ht]
    \centering
    \small
    \begin{tabular}{c|ccc|ccc|ccc|ccc|ccc}
        \textbf{RE-Adapted} & \multicolumn{3}{c|}{\textbf{Arguana}} & \multicolumn{3}{c|}{\textbf{FiQA}} & \multicolumn{3}{c|}{\textbf{NFCorpus}} & \multicolumn{3}{c|}{\textbf{SciFact}} \\
        \textbf{Model} &  w/o & w/ & both & w/o & w/ & both & w/o & w/ & both & w/o & w/ & both\\
        \toprule
        RepLLaMA & \cellcolor{green!10}+0.5 & \cellcolor{green!10}+0.2 & \cellcolor{green!10}+0.3 & \cellcolor{green!25}+1.0 & \cellcolor{green!25}+1.2 & \cellcolor{green!25}+1.3 & 
        \cellcolor{green!10}+0.3 &
        \cellcolor{green!10}+0.5 &
        \cellcolor{green!10}+0.6 &
        \cellcolor{green!10}+0.1 & \cellcolor{green!10}+0.5 & \cellcolor{green!25}+1.8 &   \\
        e5-Mistral & \cellcolor{green!50}+3.9 & \cellcolor{green!75}+5.1 & \cellcolor{green!100}+7.0 & \cellcolor{green!50}+2.0 & \cellcolor{green!50}+2.8 & \cellcolor{green!50}+2.4 & 
        \cellcolor{green!10}+0.6 &
        \cellcolor{green!25}+1.1 &
        \cellcolor{green!25}+1.2 &
        \cellcolor{green!50}+2.3 & \cellcolor{green!25}+1.0 & \cellcolor{green!25}+1.9 &  \\
    \end{tabular}
    \caption{Change in nDCG@10 over original retriever when pretrained model is fine-tuned using only documents with (w/) or without (w/o) corresponding queries in the test set, or with (both) subsets.}
    \label{tab:with_without}
\end{table*}

\paragraph{Are queried passages all that matter?} One reason the in-domain results could be better than baseline is because the passages being queried for at test-time are included in the fine-tuning data, although their corresponding queries are not. We test this hypothesis by training two additional knowledge adapters, one which sees no test-time passages, and one that sees only test-time passages. We compare the evaluation results with the original adapter fine-tuned on all the passages (\autoref{tab:with_without}). We find that neither of the subsets is always best, indicating that having in-domain data is more important than specifically fine-tuning on the passages being queried for at test-time.

\paragraph{Does any unlabeled data work?} Next, we explore the importance of using the domain specific data for fine-tuning. We repeat our main experiment across the BeIR datasets, but using only the knowledge adapter trained on MS-MARCO, our largest corpus. We compare the resulting performance with the original RepLLaMA and e5-Mistral baselines as well as the RE-Adapted models fine-tuned on the domain specific data (\autoref{tab:marco}). We observe that additional fine-tuning with MS-MARCO improves RepLLaMA by an average of 0.2 points but is still 0.5 points below the average performance when using in-domain data. For e5-Mistral however, we see that the MS-MARCO fine-tuning results in a significant increase of 2 points over baseline on average, 0.3 points above what is achieved with in-domain data. The larger gain with e5-Mistral is likely due to RepLLaMA's use of MS-MARCO as the majority of its training data, while e5-Mistral only used a subset to supplement the otherwise synthetically generated data \citep{ma2023finetuning, wang2024improving}. In both cases, the extra data improved the performance and indicates that retrieval models can generally benefit from additional unlabeled training using \readaptir{}.

\begin{table}[h]
    \centering
    \small
    \begin{tabular}{l|cc|cc}
        & \multicolumn{2}{c|}{\textbf{RepLLaMA}} & \multicolumn{2}{c}{\textbf{e5-Mistral}}\\
        \textbf{Dataset} & Base & Domn & Base & Domn \\
        \toprule
        FEVER & \cellcolor{green!10}+0.1 & \cellcolor{green!10}+0.3 & \cellcolor{red!10}-0.1 & \cellcolor{red!50}-2.7 \\
        HotPotQA & \cellcolor{red!10}-0.4 & \cellcolor{red!10}-0.8 & \cellcolor{green!10}+0.5 & \cellcolor{red!10}-0.4 \\
        NQ & \cellcolor{red!10}-0.1 & \cellcolor{red!10}-0.4 & \cellcolor{green!100}+7.0 & \cellcolor{green!100}+7.9\\ 
        Quora & \cellcolor{green!50}+2.0 & \cellcolor{red!10}-0.7 & \cellcolor{green!50}+2.4 & \cellcolor{gray!10}0.0 \\
        Arguana & \cellcolor{green!10}+0.8 & \cellcolor{green!10}+0.5 & \cellcolor{green!100}+10.1 & \cellcolor{green!50}+3.1 \\
        Climate-FEVER & \cellcolor{green!25}+1.0 & \cellcolor{green!25}+1.4 & \cellcolor{red!25}-1.1 & \cellcolor{red!75}-7.6\\
        DBPedia & \cellcolor{green!10}+0.1 & \cellcolor{gray!10}0.0 & \cellcolor{green!10}+0.5 & \cellcolor{green!10}+0.6 \\
        FiQA & \cellcolor{green!10}+0.2 & \cellcolor{red!25}-1.1 & \cellcolor{green!25}+1.7&\cellcolor{red!10}-0.7 \\
        NFCorpus & \cellcolor{gray!10}0.0 & \cellcolor{red!10}-0.6 & \cellcolor{green!10}+0.7 & \cellcolor{red!10}-0.5 \\
        SCIDOCS & \cellcolor{green!10}+0.1 & \cellcolor{red!10}-0.5 & \cellcolor{green!10}+0.6 & \cellcolor{green!10}+0.5 \\
        SciFact & \cellcolor{red!10}-0.9 & \cellcolor{red!50}-2.7 & \cellcolor{green!50}+2.1 & \cellcolor{green!10}+0.2 \\
        TREC-COVID & \cellcolor{red!10}-0.4 & \cellcolor{red!50}-2.0 & \cellcolor{green!50}+2.6 & \cellcolor{green!75}+5.5 \\
        Touche-2020 & \cellcolor{red!10}-0.1 & \cellcolor{green!10} +0.4 & \cellcolor{red!10}-0.5 & \cellcolor{red!25}-1.6\\
        \midrule
        \textbf{Average} & \cellcolor{green!10}+0.2 & \cellcolor{red!10}-0.5& \cellcolor{green!50}+2.0 & \cellcolor{green!10}+0.3 \\
    \end{tabular}
    \caption{Change in nDCG@10 when \readaptir{} is applied with pretrained model fine-tuned on MS-MARCO instead of the domain under evaluation. \textit{Base} indicates the change with respect to the original model, \textit{Domn} the change with respect to the model RE-Adapted on the evaluated domain.}
    \label{tab:marco}

\end{table} 

\section{Conclusion}

In this work, we introduced \readaptir, an extension of \readapt{} for using unlabeled data to improve the zero-shot and in-domain performance of text retrieval models. We demonstrated \readaptir{} improves two state-of-the-art models: RepLLaMA and e5-Mistral. We find that fine-tuning on the documents being queried for at test-time is not required, and still see increased performance when they are excluded. \readaptir{} improved baseline performance in two cases: one where in-domain data was used, and the other, using additional unlabeled data related to the models' original training corpus. Combined, these results enforce the wide applicability of our approach, and our findings ensure \readaptir{} is of immediate use to text-retrieval practitioners.

\bibliography{anthology, custom}

\appendix

\newpage

\section{Adapter Details}
\label{sec:details}

We trained DoRA adapters for all attention key, query, and value layers as well as the up and down projection layers. All adapters used rank 32 with alpha 64 and a LoRA dropout of 0.05.

We used a batch size of 4 using the AdamW optimizer with learning rate of 0.0002 with linear scheduling. Unless otherwise stated, all adapters were trained over 1 epoch of the respective corpus with a max length of 1024 for any example.

All training and evaluation was done using a single NVIDIA A100 GPU with 80GB of memory.

\section{Corpus Sizes}
\label{sec:sizes}

\begin{tabular}{lr}
     \textbf{Dataset} & \textbf{Corpus Size} \\
     \toprule
     MS-MARCO & 8.84M \\
     Climate-FEVER & 5.42M \\
     FEVER & 5.42M \\
     HotPotQA & 5.23M \\
     DBPedia & 4.63M \\
     NQ & 2.68M \\
     Quora & 523K \\
     Touche-2020 & 382K \\
     TREC-COVID & 171K \\
     FiQA & 57K \\
     SCIDOCS & 25K \\
     ArguAna & 8.67K \\
     SciFact & 5k \\
     NFCorpus & 3.6K \\
\end{tabular}

\end{document}